\documentclass{article}
\usepackage{amsmath,epsfig,cite}

\def\beq{\begin{eqnarray}}
\def\eeq{\end{eqnarray}}
\def\bsp{\begin{split}}
\def\esp{\end{split}}

\begin{document}

\title{\textbf{The Future of Tilted Bianchi Universes}}
\author{\textbf{John D. Barrow}\thanks{%
J.D.Barrow@damtp.cam.ac.uk} ~\textbf{and Sigbj\o rn Hervik}\thanks{%
S.Hervik@damtp.cam.ac.uk} \\
%EndAName
DAMTP, \\
Centre for Mathematical Sciences,\\
Cambridge University\\
Wilberforce Rd. \\
Cambridge CB3 0WA, UK}
\maketitle

\begin{abstract}
An asymptotic stability analysis of spatially homogeneous models of Bianchi
type containing tilted perfect fluids is performed. Using the known
attractors for the non-tilted Bianchi type universes, we check whether they
are stable against perturbations with respect to tilted perfect fluids. We
perform the analysis for all Bianchi class B models and the Bianchi type $%
VI_{0}$ model. In particular, we find that none of the non-tilted
equilibrium points are stable against tilted perfect fluids stiffer than
radiation. We also indicate parts of the phase space where new tilted exact
solutions might be found.
\end{abstract}

\section{Introduction}

In order to explain the isotropy and homogeneity of the universe and
investigate the form of its small expansion anisotropies, it is important to
understand the complete spectrum of anisotropic cosmological models. The
subset of the most general anisotropic and inhomogeneous cosmological models
that are most amenable to a full analysis are the spatially homogeneous
Bianchi universes because the Einstein field equations for these cosmologies
reduce to ordinary differential equations. They have been studied in
considerable detail by many authors but attention has largely focussed on
the behaviour of models with comoving perfect fluids. In this paper we will
extend these analyses to the situation where the fluid is 'tilted' with
respect to the normals to the hypersurfaces of homogeneity and the fluid is
not comoving.

In the study of \ Bianchi type universes there is special interest in the
particular exact solutions of the Einstein equations which are attractors
for more general classes of solution (see \cite{DynSys} for a review). The
two recent papers, \cite{HHTW,HHWVIII}, have completed the investigation of
all non-chaotic Bianchi models with a \emph{non-tilted} $\gamma $-law
perfect fluid and their asymptotic attractors are understood. This
understanding takes the form of a phase-space description of the dynamics of
the autonomous ordinary differential equations describing the cosmological
evolution. However, when we turn to Bianchi models with \emph{tilted} fluid
motion \cite{KingEllis}, the situation is far from completely understood. Up
to three additional degrees of freedom can be added to the phase spaces in
these cases. Some analyses have been made of Bianchi universes with tilted
fluids \cite{BN,BS}, in particular type $II$ \cite{HBWII} and type $V$ \cite%
{Shikin,Collins,HWV}, but a more general analysis has not been by date. The
object of this paper is to provide a local stability analysis of a wide
range of Bianchi universes containing tilted perfect fluids in order to
elucidate their late-time behaviours. Specifically, we will consider Bianchi
class B models, together with the type $VI_{0}$ model of class A, filled
with a tilted $\gamma $-law perfect fluid. We will base our analysis on the
results from the non-tilted analysis. The main reason for this is that, due
to the complexity of the equations of motion, a complete analysis would need
to find, or rule out the existence, of new tilted exact solutions which
might act as attractors. However, a relatively simple analysis based on the
results from the non-tilted models allows us to go a significant way towards
understanding the behaviour of tilted universes.

Tilted universes display many features that are of considerable interest in
the study of Einstein's equations and their physical implication. They
include rotating universes and are therefore of importance for any attempts
to formulate and evaluate Mach's Principle \cite{mach}, the conditions
needed for the appearance of closed timelike curves, and the properties of
quantum cosmologies that tunnel from 'nothing'. The changes in the
cosmological evolution of anisotropies that arise in brane-worlds are also of
special interest but, so far, the influence of non-comoving velocities have
not been studied. They are likely to be the most important factor in the
evolution as can be seen from the effects of anisotropic pressures studied
by Barrow and Maartens \cite{bm}.

Before we embark on the formal analysis of the full Einstein equations
governing the tilted Bianchi universes it is helpful to use some elementary
physical arguments to determine when we might expect critical changes in the
behaviour of velocities and vorticities to occur as the equation of state
varies. Using a fairly simple argument we can give a rough estimate of the
stability properties to be expected of non-comoving perfect fluids \cite{BS}%
. A perfect fluid with equation of state, $p=(\gamma -1)\rho $, and constant 
$\gamma $, will evolve according to $\rho \propto a^{-3\gamma }\propto
\theta ^{2}$, where $a(t)$ is the mean scale factor and $\theta =3\dot{a}/a$
is the volume expansion rate. Consider an expanding universe that is close
to isotropy: an eddy of fluid, with mass $m$, will conserve angular
momentum, $I\omega $, if 
\[
Ma^{2}\omega ={\mathrm{constant.}}
\]%
For tilted fluids with velocity $v=a\omega $ and mass $m\propto \rho a^{3},$
this implies 
\[
\rho a^{5}\omega =\rho a^{4}v={\mathrm{constant.}}
\]%
Thus for perfect fluids 
\[
v\propto a^{3\gamma -4},\quad \omega \propto a^{3\gamma -5}.
\]%
The ratio of the distorting \emph{rotational energy density} $\omega ^{2}$
to the isotropic density evolves as 
\[
\frac{\omega ^{2}}{\theta ^{2}}\propto \frac{\omega ^{2}}{\rho }\propto
a^{9\gamma -10}.
\]%
We see that we expect particular values of $\gamma $ to mark thresholds in
the evolution of models containing non-comoving velocities and vorticity. In
particular a threshold exists when $\gamma =4/3$: rotational velocities grow
with $a(t)$ for $\gamma >4/3$ \cite{turb}. The influence of vorticity on the
expansion dynamics grows when $\gamma >10/9.$ The significance of the value $%
\gamma =10/9$ can be seen, for example, in the exceptional model, discussed
section \ref{exceptional}. In the analyses to follow we find that these
critical values of $\gamma $ recur, along with the value $\gamma =2/3$,
which marks a transition between universes that isotropise at late times
(for $\gamma <2/3$) and those that do not (for $\gamma >2/3$). Of course,
the detailed behaviour is more complicated than this discussion implies
because the Bianchi types need not be close to isotropy and there are
geometrical constraints on how the different components of the 3-velocity
are able to evolve if energy and momentum are to be conserved \footnote{%
This simple physical argument can be extended to the early stages of
brane-world cosmologies. In brane-worlds we have $\theta \propto \rho $ rather
than $\theta ^{2}\propto \rho $ at early times and so the distortion to the
metric created by any vortical motions is much more significant than in
general relativistic cosmologies as the universe expands. We have $\omega
/H\propto a^{6\gamma -5}$ and rotational distortions grow in all brane-worlds
with $\gamma >5/6$ and this includes the cases of matter and radiation
domination.}.

In the next section we will determine the equations of motion for the tilted
fluid in Bianchi universes. Then, in section \ref{analysis}, we will perform
a stability analysis for the various asymptotic behaviours that are known
from the study of non-tilted universes. In section \ref{tiltedVIh} we will
analyse the type $VI_{h}$ model. Finally, in section \ref{discussion} we
summarise and discuss our results.

\section{The equations of motion}

We choose a time variable so that the time is measured along time-like
geodesics orthogonal to the surfaces of transitivity. The surfaces of
transitivity are therefore spatial and we say that the spacetime is \emph{%
spatially homogeneous}. For the Bianchi cosmologies -- which admit a simply
transitive symmetry group acting on the spatial hypersurfaces -- we can
always write the line-element as 
\[
ds^{2}=-dt^{2}+g_{ab}(t){\mbox{\boldmath${\omega}$}}^{a}{\mbox{\boldmath${%
\omega}$}}^{b},
\]%
where $g_{ab}$ is a non-singular symmetric $3\times 3$ matrix only dependent
upon time, ${\mbox{\boldmath${\omega}$}}^{a}$ is a triad of one-forms
obeying 
\[
\mathbf{d}{\mbox{\boldmath${\omega}$}}^{a}=-\frac{1}{2}C_{~bc}^{a}{%
\mbox{\boldmath${\omega}$}}^{b}\wedge {\mbox{\boldmath${\omega}$}}^{c},
\]%
and $C_{~bc}^{a}$ are the structure constants of the Bianchi group type
under consideration. The structure constants $C_{~bc}^{a}$ can be split into
a vector part $a_{b}$, and a trace-free part $n^{ab}$ by \cite{EM} 
\[
C_{~bc}^{a}=\varepsilon _{bcd}n^{da}-\delta _{~b}^{a}a_{c}+\delta
_{~c}^{a}a_{b}.
\]%
The matrix $n^{ab}$ is symmetric, and, using the Jacobi identity, $%
a_{b}=(1/2)C_{~ba}^{a}$ is in the kernel of $n^{ab}$ 
\[
n^{ab}a_{b}=0.
\]

The Bianchi types are divided into two classes:

\begin{itemize}
\item {} Class A: $a_c=0$.

\item {} Class B: $a_c\neq 0$.
\end{itemize}

First, we will concentrate on the class B spacetimes. In this case we can
choose the orientation of the triad so that $a_{c}=a\delta _{~c}^{1}$. This
implies that $n^{ab}$ is of rank 2 and can be written 
\[
\left( n^{ab}\right) =%
\begin{bmatrix}
0 & 0 & 0 \\ 
0 & n^{22} & n^{23} \\ 
0 & n^{23} & n^{33}%
\end{bmatrix}%
.
\]%
Consider a perfect fluid with energy-momentum tensor 
\[
T_{\mu \nu }=(\rho +p)u_{\mu }u_{\nu }+pg_{\mu \nu }.
\]%
The normalised 4-velocity of the fluid $u^{\mu }$ is in general tilted with
respect to the normals to the hypersurfaces of homogeneity, $n_{a}=t_{_{,}a}$%
; and the fluid three-velocity $v^{a}$, where $u^{\mu }=(u^{0},v^{a})$, will
be assumed to be non-zero. The equations of motion for the fluid, $T_{~~;\nu
}^{\mu \nu }=0$, can be written as a closed system of equations \cite%
{DG,Dem1} 
\begin{eqnarray}
-(\rho +p)\dot{v}_{a}v^{a} &=&\dot{p}v_{a}v^{a}  \label{eq:one} \\
-(\rho +p)\left( u^{0}\dot{v}_{c}+v_{a}v^{b}C_{~bc}^{a}\right)  &=&\dot{p}%
u^{0}v_{c}  \label{eq:two} \\
(\rho +p)\left( \dot{u}^{0}+v^{b}a_{b}+u^{0}\theta \right) +u^{0}\dot{\rho}
&=&0.  \label{eq:three}
\end{eqnarray}%
Here, $\theta $ is the volume expansion factor $\theta =\dot{g}/2g$. In
our analysis, eqs. (\ref{eq:one}-\ref{eq:three}) are the essential equations
because we will consider the stability of non-tilted solutions, that obey
the Einstein equations with $v_{a}\equiv 0,$ with respect to the
perturbations introduced by the present of non-comoving velocities.

Recently, the two remaining Bianchi models with non-tilted $\gamma $-law
perfect fluid have been analysed \cite{HHTW,HHWVIII}; so all non-tilted
Bianchi models have now been analysed and their asymptotic behaviours
determined. Using these results, which give all the attractors in the set of
non-tilted perfect fluid cosmologies, we can investigate the stability of
the late-time non-tilted attractors against perturbations from tilted matter.

It is useful to introduce a orthonormal triad $\mathsf{e}_{~a}^{\hat{a}}(t)$
by defining 
\[
\delta _{\hat{a}\hat{b}}\mathsf{e}_{~a}^{\hat{a}}\mathsf{e}_{~b}^{\hat{b}%
}=g_{ab},\quad \delta _{\hat{a}\hat{b}}=\mathsf{e}_{~\hat{a}}^{a}\mathsf{e}%
_{~\hat{b}}^{b}g_{ab}.
\]%
We can now write eqs. (\ref{eq:one}-\ref{eq:three}) in terms of the
components $v_{\hat{a}}=\mathsf{e}_{~\hat{a}}^{a}v_{a}$ in the orthonormal
frame: 
\begin{eqnarray}
-(\rho +p)\left( \dot{v}_{\hat{a}}v^{\hat{a}}+\dot{\mathsf{e}}_{~a}^{\hat{a}}%
\mathsf{e}_{~\hat{b}}^{a}v_{\hat{a}}v^{\hat{b}}\right) 
&=&\dot{p}v_{\hat{a}}v^{\hat{a}} \label{eq:one2} \\
-(\rho +p)\left( u^{0}\dot{v}_{\hat{c}}+u^{0}\mathsf{e}_{~\hat{c}}^{a}\dot{%
\mathsf{e}}_{~a}^{\hat{b}}v_{\hat{b}}+v_{\hat{a}}v^{\hat{b}}\mathsf{e}_{~a}^{%
\hat{a}}\mathsf{e}_{~\hat{b}}^{b}C_{~bc}^{a}\mathsf{e}_{~\hat{c}}^{c}\right)
&= &\dot{p}u^{0}v_{\hat{c}}  \label{eq:two2} \\
(\rho +p)\left( \dot{u}^{0}+v^{\hat{b}}\mathsf{e}_{~\hat{b}%
}^{b}a_{b}+u^{0}\theta \right) +u^{0}\dot{\rho}&=&0 .  \label{eq:three2}
\end{eqnarray}%
The stability of a particular solution can now be checked by linearising
these equations around the exact non-tilted solution describing the
asymptote in question. For non-tilted equilibrium points, the linearised
version of eqs. (\ref{eq:one2}-\ref{eq:two2}) will be a closed set of
equations, and hence, the stability analysis can to a large extent be done
by considering these equations alone (provided the stability analysis has
already been performed for the remaining equations).

More precisely, our stability analysis is based on the following
observation. The total dynamical system (including the Einstein equations)
can be written close to an equilibrium point, say at $\mathbf{x}=0$, 
\[
\dot{\mathbf{x}}=\mathsf{A}\mathbf{x}+\mathbf{F}(\mathbf{x}),
\]%
where $\mathbf{F}(\mathbf{x})$ is at least quadratic in $\mathbf{x}$. In our
case, when considering a non-tilted equilibrium point, the matrix $\mathsf{A}
$ will be of the block form 
\begin{eqnarray}
\mathsf{A}=%
\begin{bmatrix}
\mathsf{A_{1}} & \mathsf{0} \\ 
\mathsf{B} & \mathsf{A}_{2}%
\end{bmatrix}%
.  \label{eq:Amatrix}
\end{eqnarray}
This means that the eigenvalues of $\mathsf{A}$ are the union of the
eigenvalues of $\mathsf{A}_{1}$ and $\mathsf{A}_{2}$. Thus if we choose
equilibrium points that are stable in the set of non-tilted cosmologies, we
only have to check the eigenvalues for $\mathsf{A}_{1}$; i.e. the linearised
version of eq. (\ref{eq:two2}). This simplifies the stability analysis
considerably and turns out to be very fruitful. However, there are some limits
to this procedure:

\begin{enumerate}
\item {} If the real parts of one or more eigenvalues of $\mathsf{A}_{1}$
are zero, then the stability analysis is inconclusive from analysing $%
\mathsf{A}_{1}$ alone.

\item {} For tilted equilibrium points, $\mathsf{A}$ will in general not
have the form (\ref{eq:Amatrix}); i.e. the complete set of equations is
needed to establish its stability.
\end{enumerate}

Keeping these limitations in mind we shall perform a perturbation analysis
for various spacetimes.

\section{Stability analysis}

\label{analysis}We will study the various class B cases, plus the type $%
VI_{0}$ case which is closely related. We will henceforth assume that the
tilted perfect fluid obeys the barotropic equation of state\footnote{%
Non-perfect fluid stresses, for example anisotropic pressures, can have
significant effects on the evolution of anisotropies; see for example refs. 
\cite{jb,bm}.} 
\[
p=(\gamma -1)\rho .
\]%
Note that ever-expanding cosmologies with inflationary types of fluids with $%
0\leq \gamma <2/3$, will all asymptote to a FRW model with $a(t)\propto
t^{2/3\gamma }$ as $t\rightarrow \infty $ in accord with the cosmological
\textquotedblleft no-hair\textquotedblright\ theorem \cite{Wald,DynSys}.
These fluids will therefore not be considered explicitly in what follows.
However, the $\gamma =2/3$ will appear as a stability boundary for the
anisotropic attractors as expected.

The notion of tilt does not apply to an exact vacuum solution.  Hence,
for equilibrium points corresponding to vacuum spacetimes we can only
define a tilt when they are perturbed away from any exact vacuum
solutions.  Thus, in what follows, when refer to stability with respect
to the introduction of tilt we are determining whether the universe at
late times is best described by a tilted or a non-tilted model. Also,
when referring to the vacuum solutions we will always mean the
non-tilted equilibrium points. 

\subsection{Bianchi type $VII_{h}$}

The most general Bianchi universes containing the open Friedmann universe as
a subcase are those of type $VII_{h}$. The general late-time asymptotes for
non-tilted spacetimes of type $VII_{h}$ are the plane-wave vacuum solutions
of Lukash \cite{Lukash}.\footnote{%
These vacuum solutions can easily be generalised to plane-wave solutions
with an electromagnetic field \cite{sigPP}.} For these solutions \cite{sigPP}
\begin{eqnarray}
{\mbox{\boldmath${\omega}$}}^{1} &=&\mathbf{dx} \\
{\mbox{\boldmath${\omega}$}}^{2} &=&e^{sx}\left[ \cos \omega x\mathbf{dy}%
-\sin \omega x\mathbf{dz}\right]  \\
{\mbox{\boldmath${\omega}$}}^{3} &=&e^{sx}\left[ \sin \omega x\mathbf{dy}%
+\cos \omega x\mathbf{dz}\right]  \\
s(1-s) &=&\omega ^{2}\sinh ^{2}2\beta ,
\end{eqnarray}%
and 
\[
\mathsf{e}_{~a}^{\hat{a}}=%
\begin{bmatrix}
t & 0 & 0 \\ 
0 & t^{s}e^{\beta }\cos (\omega \ln t) & -t^{s}e^{\beta }\sin (\omega \ln t)
\\ 
0 & t^{s}e^{-\beta }\sin (\omega \ln t) & t^{s}e^{-\beta }\cos (\omega \ln t)%
\end{bmatrix}%
\]%
The group parameter, $h$, is given by 
\[
h=\frac{s^{2}}{\omega ^{2}}.
\]%
Furthermore, $\theta =(2s+1)/t,$ and they reduce to isotropy when $s=1$ and $%
\omega =\beta =0$. Eq. (\ref{eq:three2}) is to lowest order in $v_{\hat{a}}$ 
\beq
\dot{\rho}+\gamma \theta \rho =0.
\label{eq:rhodot}\eeq
This is the same as for non-tilted fluid. Defining expansion-normalised
energy-density by $\Omega =\rho /\theta ^{2}$, we require $\dot{\Omega}<0$
for the solutions to be future stable. This yields the bound 
\[
\frac{2}{1+2s}<\gamma .
\]%
This is the usual bound required for the stability of these solutions in the
space of non-tilted $VII_{h}$ universes with perfect fluid found earlier 
\cite{DynSys,BS1,BS}. Linearising eq. (\ref{eq:two2}), using eq. (\ref%
{eq:rhodot}), and dividing by $\gamma \rho $, we get the following system of
equations for the 3-velocities: 
\begin{eqnarray}
{v}_{\hat{1}}^{\prime } &=&\left[ (2s+1)\gamma -(2s+2)\right] {v}_{\hat{1}} 
\notag \\
{v}_{\hat{2}}^{\prime } &=&\left[ (2s+1)\gamma -(3s+1)\right] {v}_{\hat{2}%
}-e^{2\beta }\omega {v}_{\hat{3}}  \notag \\
{v}_{\hat{3}}^{\prime } &=&\left[ (2s+1)\gamma -(3s+1)\right] {v}_{\hat{3}%
}+e^{-2\beta }\omega {v}_{\hat{2}}.
\end{eqnarray}%
Here, we have also introduced a new time-variable $\tau =\ln t$ and a prime
denotes derivative with respect to $\tau $. The eigenvalues of this system
of equations, which determine the stability of the plane-wave solutions, are 
\begin{eqnarray}
\lambda _{1} &=&(2s+1)\gamma -(2s+2),  \notag \\
\lambda _{2,3} &=&(2s+1)\gamma -(3s+1)\pm i\omega .
\end{eqnarray}%
Hence, the plane-wave solutions of type $VII_{h}$ are stable when 
\[
\frac{2}{2s+1}<\gamma <\frac{3s+1}{2s+1}.
\]%
Note that $(3s+1)/(2s+1)\leq 4/3$; so the solutions are unstable for fluids
stiffer than radiation. This reflects the tendency for very stiff fluids to
'spin up' as they expand in order to conserve angular momentum \cite{turb},
as discussed above.

For radiation, $\gamma =4/3$, the solutions are unstable for $s<1$. The case 
$s=1$ yields a zero eigenvalue for radiation and has to be considered
separately. However, $s=1$ implies that the solutions are LRS and are
therefore the same as the LRS type $V$ solutions. Type $V$ evolution has
been considered elsewhere, and, in particular, the stability of this
solution has been analysed in detail \cite{Shikin,Collins,HWV}. For \emph{%
generic} perturbations\footnote{%
In \cite{HWV}, it is shown that this solution is, in fact, an attractor for
a set of non-zero measure of solution trajectories in the state space of
irrotational type V models. However, for some perturbations it is unstable
and is therefore generically unstable.} this solution is found to be \emph{%
unstable}. Hence, all plane-wave solutions of type $VII_{h}$ are unstable
for radiation ($\gamma =4/3$).

Since for a given $s$, any value of $h$ is possible, we can conclude that
for any given $h$ there are plane-wave solutions of type $VII_{h}$ that are
stable in the set of $VII_{h}$ models, whenever 
\[
\frac{2}{3}<\gamma <\frac{4}{3}.
\]%
A summary of the results for type $VII_{h}$ is illustrated in Fig. \ref%
{fig:typeVIIh}. 
\begin{figure}[tbp]
\centering \epsfig{figure=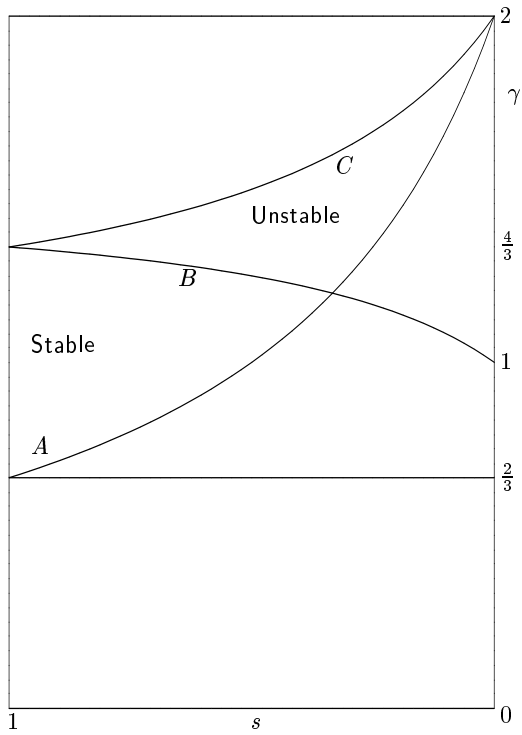, width=7cm}
\caption{Stability diagram for the Bianchi type VII$_{h}$ Lukash plane-wave
solutions. Here, the labelled curves are defined by the following functions: 
$A~$is $\protect\gamma =\frac{2}{2s+1}$, $B$ is$~\protect\gamma =\frac{3s+1}{%
2s+1}$ and $C$ is$~\protect\gamma =\frac{2s+2}{2s+1}$. The region marked
\textquotedblleft \textsf{Stable}\textquotedblright\ is bounded by the
curves $A$ and $B$, while the region marked \textquotedblleft \textsf{%
Unstable}\textquotedblright\ is bounded by $A$, $B$ and $C$ as shown. The
latter region has two directions in which the solutions are unstable against
tilt. All the remaining unmarked regions of the diagram are also those in
which the Lukash solutions are unstable.}
\label{fig:typeVIIh}
\end{figure}

\subsection{Bianchi type $V$}

The general late-time attractor for non-tilted type $V$ models is the
isotropic Milne universe. This vacuum solution can be extracted from the
Lukash plane waves above by choosing $\omega =0$, $s=1$. Hence, from the
above we see that this solution is stable provided that 
\[
\frac{2}{3}<\gamma <\frac{4}{3}.
\]%
This agrees with earlier results of other authors \cite%
{BS,Shikin,Collins,HWV}.

\subsection{Bianchi type $VI_{h}$}

Consider the type $VI_{h}$ plane-wave solutions given by 
\begin{eqnarray}
{\mbox{\boldmath${\omega}$}}^{1} &=&\mathbf{dx} \\
{\mbox{\boldmath${\omega}$}}^{2} &=&e^{(s+b)x}\mathbf{dy} \\
{\mbox{\boldmath${\omega}$}}^{3} &=&e^{(s-b)x}\mathbf{dz} \\
s(1-s) &=&b^{2}+r^{2},
\end{eqnarray}%
and 
\[
\mathsf{e}_{~a}^{\hat{a}}=%
\begin{bmatrix}
t & 0 & 0 \\ 
0 & t^{s+b} & -\frac{r}{b}t^{s-b} \\ 
0 & 0 & t^{s-b}%
\end{bmatrix}%
.
\]%
The group parameter is given by 
\[
h=-\frac{s^{2}}{b^{2}},
\]%
and $\theta =(2s+1)/t$. Following the same procedure as in the type $VII_{h}$
case, we obtain a lower bound from eq. (\ref{eq:three2}): 
\beq
\frac{2}{1+2s}<\gamma .
\label{eq:lowerbound}\eeq
Linearising eq. (\ref{eq:two2}) we get 
\begin{eqnarray}
{v}_{\hat{1}}^{\prime } &=&\left[ (2s+1)\gamma -(2s+2)\right] {v}_{\hat{1}},
\notag \\
{v}_{\hat{2}}^{\prime } &=&\left[ (2s+1)\gamma -(3s+1+b)\right] {v}_{\hat{2}%
},  \notag \\
{v}_{\hat{3}}^{\prime } &=&\left[ (2s+1)\gamma -(3s+1-b)\right] {v}_{\hat{3}%
},
\end{eqnarray}%
and the eigenvalues are: 
\begin{eqnarray}
\lambda _{1} &=&(2s+1)\gamma -(2s+2),  \notag \\
\lambda _{2,3} &=&(2s+1)\gamma -(3s+1\pm b).
\end{eqnarray}%
Assuming that bound (\ref{eq:lowerbound}) holds, we have 
\[
\lambda _{3}>2-(3s+1-b)\geq \frac{3}{2}(1-\sqrt{-h}).
\]%
Hence, all models with $|h|<1$ are unstable. The threshold model, of LRS
type $III=VI_{-1}$, has one solution with a zero eigenvalue.

Combining these results, we see that for $|h|>1$ there are stable plane wave
solutions whenever 
\[
\frac{2(1-h)}{1-3h}<\gamma <\frac{1-\sqrt{-h}-4h}{1-3h}<\frac{4}{3}.
\]

Consider next the Collins $VI_{h}$ perfect fluid solution 
\begin{eqnarray}
{\mbox{\boldmath${\omega}$}}^{1} &=&\mathbf{dx} \\
{\mbox{\boldmath${\omega}$}}^{2} &=&e^{\frac{r(2-\gamma )+c}{2\gamma }x}%
\mathbf{dy} \\
{\mbox{\boldmath${\omega}$}}^{3} &=&e^{\frac{r(2-\gamma )-c}{2\gamma }x}%
\mathbf{dz} \\
c^{2} &=&(2-\gamma )(3\gamma -2),
\end{eqnarray}%
and 
\[
\mathsf{e}_{~a}^{\hat{a}}=%
\begin{bmatrix}
t & 0 & 0 \\ 
0 & t^{\frac{2-\gamma +rc}{2\gamma }} & 0 \\ 
0 & 0 & t^{\frac{2-\gamma -rc}{2\gamma }}%
\end{bmatrix}%
,
\]%
where $0<r<1$, and 
\[
h=-\frac{(2-\gamma )^{2}r^{2}}{c^{2}}.
\]%
This solution is stable to the future for non-tilted perfect-fluid
cosmologies whenever 
\beq
\frac{2}{3}<\gamma <\frac{2(1-h)}{1-3h}.
\label{eq:boundVIh2}\eeq
Linearising eq. (\ref{eq:two2}), we get 
\begin{eqnarray}
{v}_{\hat{1}}^{\prime } &=&\frac{\gamma -2}{\gamma }{v}_{\hat{1}}  \notag \\
{v}_{\hat{2}}^{\prime } &=&\frac{1}{2\gamma }\left( 5\gamma -6-rc\right) {v}%
_{\hat{2}}  \notag \\
{v}_{\hat{3}}^{\prime } &=&\frac{1}{2\gamma }\left( 5\gamma -6+rc\right) {v}%
_{\hat{3}}.
\end{eqnarray}%
We note that 
\[
\lambda _{1}<0,\quad \lambda _{2}<\lambda _{3}
\]%
for $\gamma <2$ and so stability is assured if 
\[
\lambda _{3}<0\Rightarrow -h<\frac{(5\gamma -6)^{2}}{(3\gamma -2)^{2}};
\]%
or in terms of $\gamma $, 
\[
\gamma <\frac{2(3+\sqrt{-h})}{5+3\sqrt{-h}}.
\]%
When $-h<1$, this bound in tighter than the upper bound in eq. (\ref%
{eq:boundVIh2}); when $-h>1$ the bound (\ref{eq:boundVIh2}) is the tightest.
Hence, we can conclude that there are future stable Collins type $VI_{h}$
perfect fluid solutions in the presence of tilt so long as 
\[
\frac{2}{3}<\gamma <\min \left( \frac{2(1-h)}{1-3h},\frac{2(3+\sqrt{-h})}{5+3%
\sqrt{-h}}\right) <\frac{6}{5}.
\]

\subsection{Bianchi type $IV$}

The late-time attractors for non-tilted type $IV$ models are also plane
waves. Their stability can be obtained from the type $VI_{h}$ analysis by
taking the limit $b\rightarrow 0$ to obtain the eigenvalues 
\begin{eqnarray}
\lambda _{1} &=&(2s+1)\gamma -(2s+2),  \notag \\
\lambda _{2,3} &=&(2s+1)\gamma -(3s+1).
\end{eqnarray}%
Hence, there will be stable solutions whenever 
\[
\frac{2}{3}<\gamma <\frac{4}{3}.
\]

\subsection{The exceptional model of Bianchi type $VI_{-1/9}^{\ast }$}

\label{exceptional} The asymptotic dynamics of the exceptional model has
been studied with a non-tilted perfect fluid by Hewitt \cite{Hewitt}, and
more recently by Horwood \textit{et al} \cite{HHTW}. A stability analysis
was done for both non-tilted fluid and with fluids with a single non-zero
tilted component by Barrow and Sonoda \cite{BS}.

Consider the Robinson-Trautman vacuum solution 
\begin{eqnarray}
{\mbox{\boldmath${\omega}$}}^{1} &=&\mathbf{dx} \\
{\mbox{\boldmath${\omega}$}}^{2} &=&e^{\frac{\sqrt{6}}{5}x}\mathbf{dy} \\
{\mbox{\boldmath${\omega}$}}^{3} &=&e^{-\frac{2\sqrt{6}}{5}x}\mathbf{dz}
\end{eqnarray}%
and 
\[
\mathsf{e}_{~a}^{\hat{a}}=%
\begin{bmatrix}
t & 0 & 0 \\ 
\frac{\sqrt{5}}{2}t & t^{\frac{1}{5}} & 0 \\ 
0 & 0 & t^{\frac{3}{5}}%
\end{bmatrix}%
.
\]%
This solution has unusual volume expansion, with $\theta =9/(5t)$. This
solution is an attractor for all non-tilted perfect fluids with $10/9\leq
\gamma $. This requirement for stability is easily seen to come from eq. (%
\ref{eq:rhodot}). Linearising eq. (\ref{eq:two2}), we obtain 
\begin{eqnarray}
{v}_{\hat{1}}^{\prime } &=&\frac{1}{5}(9\gamma -14){v}_{\hat{1}}  \notag \\
{v}_{\hat{2}}^{\prime } &=&\frac{1}{5}(9\gamma -10){v}_{\hat{2}}  \notag \\
{v}_{\hat{3}}^{\prime } &=&\frac{3}{5}(3\gamma -4){v}_{\hat{3}}.
\end{eqnarray}%
However, due to the exact vanishing of the $02$-equation of Einstein's field
equations, we have to set $v_{\hat{2}}=0$. Hence, the type $VI_{-1/9}^{\ast }
$ implies that only $v_{\hat{1}}$ and $v_{\hat{3}}$ can be non-zero. Thus we
have only two eigenvalues to consider 
\[
\lambda _{1}=\frac{1}{5}(9\gamma -14),\quad \lambda _{3}=\frac{3}{5}(3\gamma
-4).
\]%
The first of these eigenvalues was found by Barrow and Sonoda \cite{BS}.
However, $\lambda _{3}$ gives an even tighter bound for stability. Thus, the
Robinson-Trautman solution is stable if 
\[
\frac{10}{9}\leq \gamma <\frac{4}{3}.
\]

For $\gamma =10/9$, the attractor for the non-tilted case is the Wainwright $%
VI_{-1/9}^{\ast }$ $\gamma =10/9$ perfect fluid solution \cite%
{Wainwright10-9} which interpolates between the Collins perfect-fluid
solutions, studied earlier, and the Robinson-Trautman solution. The solution
is 
\begin{eqnarray}
{\mbox{\boldmath${\omega}$}}^{1} &=&\mathbf{dx} \\
{\mbox{\boldmath${\omega}$}}^{2} &=&e^{\frac{r\sqrt{6}}{5}x}\mathbf{dy} \\
{\mbox{\boldmath${\omega}$}}^{3} &=&e^{-\frac{2r\sqrt{6}}{5}x}\mathbf{dz}
\end{eqnarray}%
and 
\[
\mathsf{e}_{~a}^{\hat{a}}=%
\begin{bmatrix}
t & 0 & 0 \\ 
wt & t^{\frac{1}{5}} & 0 \\ 
0 & 0 & t^{\frac{3}{5}}%
\end{bmatrix}%
.
\]%
Here, we have $\theta =9/(5t)$, and $w^{2}=9r^{2}/4-1$.

Linearising eq. (\ref{eq:two2}), we get the eigenvalues 
\[
\lambda _{1}=-\frac{4}{5},\quad \lambda _{3}=-\frac{2}{5}.
\]%
Since both are negative, this solution set is asymptotically stable as well.

When $\gamma <10/9$, the non-tilted solutions asymptote to the Collins $%
VI_{-1/9}$ perfect-fluid solution studied earlier. However, here we must
again exclude one of the eigenvalues because of the vanishing of the $(02)$
Einstein equation. Nonetheless, all eigenvalues are negative when $h=-1/9$
and $\gamma <10/9$, so this does not tighten the bounds already arising in
the non-tilted analysis. Hence, there are stable solutions of the
exceptional model when 
\[
\frac{2}{3}<\gamma <\frac{10}{9}.
\]

\subsection{Bianchi type $VI_{0}$}

This model can be obtained as the $h\rightarrow 0$ limit of type $VI_{h}$
universes, both geometrically and dynamically. In the non-tilted case, the
Collins $VI_{0}$ perfect-fluid solution is the future attractor for $%
2/3<\gamma <2$. Allowing for tilted fluids we see that the Collins $VI_{0}$
perfect-fluid solution is stable whenever 
\[
\frac{2}{3}<\gamma <\frac{6}{5}.
\]

\begin{figure}[tbp]
\centering \epsfig{figure=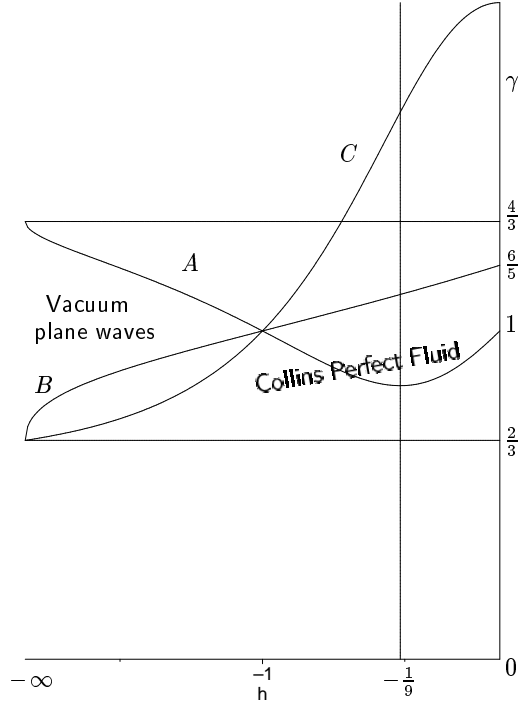, width=7cm}
\caption{Stability diagram showing the stable solutions for the Bianchi type 
$VI_{h}$ cases. The labelled curves are defined by the following functions: $%
A$ is$~\protect\gamma =\frac{1-\protect\sqrt{-h}-4h}{1-3h}$, $B$ is$~\protect%
\gamma =\frac{2(3+\protect\sqrt{-h})}{5+3\protect\sqrt{-h}}$ and $C$ is$~%
\protect\gamma =\frac{2(1-h)}{1-3h}$. The region marked \textsf{%
\textquotedblleft Vacuum plane waves\textquotedblright }\ is bounded by the
curves $A$ and $C$, while the region marked \textquotedblleft \textsf{%
Collins perfect fluid}\textquotedblright\ is bounded by $\protect\gamma =2/3$%
, $B$ and $C$. The region bounded by $\protect\gamma =4/3$, $A$ and $B$
contains none of the solutions we have investigated as attractors, except
for the exceptional case which cuts through this region at $h=-1/9$. For
non-tilted cosmologies the region corresponding to \textquotedblleft \textsf{%
Collins perfect fluid}\textquotedblright\ is the entire region bounded by $%
\protect\gamma =2/3$ and $C$; while for \textquotedblleft \textsf{Vacuum
plane waves}\textquotedblright\ it is the region above $C$.}
\label{fig:typeVI}
\end{figure}

\section{Tilted Bianchi type $VI_{\lowercase{h}}$ equilibrium points}

\label{tiltedVIh} An interesting point emerges from the analysis of the type 
$VI_{h}$ models. We can see that the instability of the Collins
perfect-fluid solution in the exceptional model with $\gamma =10/9$
corresponds to an instability of the Collins solutions with respect to a
tilted fluid for more general $h$ (see Fig. \ref{fig:typeVI}, where this is
illustrated by the fact that the line $B$ intersects $h=-1/9$ at $\gamma
=10/9$). Hence, at $h=-1/9$, the instability with respect to tilt is
exchanged for an instability of the shear: when $h=-1/9$ there does not need
to be a matter source to drive the solution away from the Collins solutions.
This may indicate that the exceptional model, in the presence of tilted
fluid, is not as dynamically exceptional as one might think and that the
rest of the type $VI_{h}$ models will behave in a similar way. Thus there
should exist exact tilted solutions of type $VI_{h}$, analogous to the
Robinson-Trautman vacuum solutions, which act as attractors. Such solutions
are not known in general, but we expect them to exist.

Hence, one is tempted to search for solutions on the form 
\begin{eqnarray}
{\mbox{\boldmath${\omega}$}}^{1} &=&\mathbf{dx} \\
{\mbox{\boldmath${\omega}$}}^{2} &=&e^{(s-b)x}\mathbf{dy} \\
{\mbox{\boldmath${\omega}$}}^{3} &=&e^{(s+b)x}\mathbf{dz}
\end{eqnarray}%
and 
\[
\mathsf{e}_{~a}^{\hat{a}}=%
\begin{bmatrix}
t & 0 & 0 \\ 
w_{1}t & t^{\sigma } & 0 \\ 
w_{2}t & 0 & t^{\tau }%
\end{bmatrix}%
,
\]%
where $w_{1,2},\sigma ,\tau ,s$ and $b$ are constants. Due to the power-law
dependence of the triad, this metric is self-similar by construction and the
Einstein equations reduce to purely algebraic equations. However, the
equations are still quite complex and difficult to solve in full generality.
Rosquist \cite{Rosquist}, and Rosquist and Jantzen \cite{RJ} have found some
solutions of Bianchi type $VI_{0}$ of this type. In general, such spacetimes
will describe solutions with tilt and non-zero vorticity. Furthermore, due
to the complexity of the solutions no analysis of their stability has been
done. Preliminary results indicate that the solutions found by Rosquist and
Jantzen may play an important role in the late-time behaviour for at least a
class of tilted type $VI_{0}$ models \cite{typeVI0}. Notwithstanding the
lack of a complete analysis, we believe that these kinds of solutions may be
important for the late-time behaviour of models in the remaining region in
Fig. \ref{fig:typeVI}; namely that bounded by the curves $\gamma =4/3$, $A,$
and $B$. One piece of supporting evidence for this conjecture is, as already
mentioned, the fact that the exceptional case intersects this region. We
believe that the Robinson-Trautman solution shares many common features with
these \textquotedblleft missing\textquotedblright\ solutions (apart from the
fact that this is the only vacuum solution).

\section{Discussion}

\label{discussion} A summary of our results are given in table \ref{table}.

We have studied the stability of all the exact non-tilted class B and type $%
VI_{0}$ Bianchi-type universes which act as attractors for the general
non-tilted cosmological evolution at late times. This has led to a
determination of their local stability in the presence of non-comoving
motions fluid motions. None of the non-tilted attractors are stable with
respect to perfect fluids stiffer that radiation ($p>\rho /3$). The type $%
VI_{h}$ models show a fairly complex dependence on the equation of state and
the group parameter defining the 3-curvature anisotropy of the universe.
However, there may be indications from this analysis that there are some
\textquotedblleft missing\textquotedblright\ stable models of this type
which have not yet been found. The exceptional $VI_{-1/9}^{\ast }$ model is
the only case with known exact solutions of this type. Further analysis of
this \textquotedblleft missing\textquotedblright\ set of solutions is
required. For type $VII_{h}$, the situation is simpler: whenever $2/3<\gamma
<4/3$ there are always some Lukash plane-wave metrics that are stable for
any given $h\neq 0$.

This investigation has been a step towards acquiring a complete
understanding of all tilted Bianchi models. Clearly, there are some missing
pieces in the analysis: class A universes require a separate analysis and
amongst those of class B the most prominent lacuna is the \textquotedblleft
missing\textquotedblright\ set of solutions of type $VI_{h}$. Also, we have
only done a local stability analysis of non-tilted equilibrium point. The
type $V$ analysis \cite{HWV} shows that the evolution in the whole state
space may be far more complex than local perturbative analyses can reveal.
New equilibrium points with non-zero tilt could exist which act as
additional attractors. Furthermore, due to the nature of our method, the
stability of the solutions along the boundary between stable and unstable
regions (i.e. those that have one or more eigenvalues with zero real parts)
remains unsettled. These models require further analysis. The Bianchi type $%
III$ universes appear interesting in this context. The type $III$ dust model
($\gamma =1$) is the centre model in Fig. \ref{fig:typeVI} where the curves $%
A$, $B$ and $C$ all intersect. Hence, the neighbouring models appear to have
a very sensitive dependence on the defining parameters.

This study is has been a first attempt to extend our systematic
understanding of the dynamics of Bianchi type universes to those with tilted
fluid motions. It should be viewed as a first step towards gaining that
understanding. We hope that the lacunae we have defined and the stability
analyses we have presented will provide a basis for the development of a
complete picture of the behaviour of tilted Bianchi type universes. Although
non-comoving fluid motions have generally been neglected in studies of
anisotropic models of the early universe, they are by no means
negligible \cite{UEWE}.
Under a wide range of realistic equations of state they have been shown to
be the most important source of anisotropy in expanding universes. In the
case of brane-world cosmologies \cite{bm2} the effects of these motions will
be even more dominant and no reliable understanding of the early evolution
of these cosmologies will be possible in general without a full analysis of
tilted fluid motions and vorticity. 
\begin{table}[tbp]
\centering 
\begin{tabular}{|c|c|c|}
\hline
Solution Type & Matter & Stability \\ \hline\hline
VII$_{h}$ (Lukash) & $\frac{2}{3}<\gamma <\frac{4}{3}$ & Stable \\ 
& $\frac{4}{3}\leq \gamma \leq 2$ & Unstable \\ \hline
IV (p-wave) & $\frac{2}{3}<\gamma <\frac{4}{3}$ & Stable \\ 
& $\frac{4}{3}\leq \gamma \leq 2$ & Unstable \\ \hline
V (Milne) & $\frac{2}{3}<\gamma <\frac{4}{3}$ & Stable \\ 
& $\frac{4}{3}\leq \gamma \leq 2$ & Unstable \\ \hline
VI$_{h}$ (Collins) & $\frac{2}{3}<\gamma <\min \left( \frac{2(1-h)}{1-3h},%
\frac{2(3+\sqrt{-h})}{5+3\sqrt{-h}}\right) $ & Stable \\ 
& $\frac{2(3+\sqrt{-h})}{5+3\sqrt{-h}}<\gamma <\frac{2(1-h)}{1-3h}$ & 
Unstable \\ \hline
VI$_{h}$ (p-wave) & $\frac{2(1-h)}{1-3h}<\gamma <\frac{1-\sqrt{-h}-4h}{1-3h}$
& Stable \\ 
& $\frac{1-\sqrt{-h}-4h}{1-3h}<\gamma \leq 2$ & Unstable \\ \hline
VI$_{-1/9}^{\ast }$ (Collins) & $\frac{2}{3}<\gamma <\frac{10}{9}$ & Stable
\\ \hline
VI$_{-1/9}^{\ast }$ (W) & $\gamma =\frac{10}{9}$ & Stable \\ \hline
VI$_{-1/9}^{\ast }$ (R-T) & $\frac{10}{9}<\gamma <\frac{4}{3}$ & Stable \\ 
& $\frac{4}{3}<\gamma \leq 2$ & Unstable \\ \hline
VI$_{0}$ (Collins) & $\frac{2}{3}<\gamma <\frac{6}{5}$ & Stable \\ 
& $\frac{6}{5}<\gamma \leq 2$ & Unstable \\ \hline
\end{tabular}%
\caption{A summary of our results: the exact solutions listed in the first
column are all attractors in the non-tilted case for the equation of state
and model parameters indicated in second column. The third column records
whether they are stable or unstable when tilted fluids are introduced. Here,
W=Wainwright, and R-T=Robinson-Trautman.}
\label{table}
\end{table}

\section*{Acknowledgments}

SH was funded by the Research Council of Norway and an Isaac Newton
Studentship.

\end{document}